# Investigating Organic Carbon and Thermal History of CM Carbonaceous Chondrites Using Spectroscopy and Laboratory Techniques

Safoura. Tanbakouei[1], [1]Department of Earth Science, The University of Hong Kong, Hong Kong, China, stanba@hku.hk, Corresponding author
Rui-Lin Cheng[1], [1]Department of Earth Science, The University of Hong Kong, Hong Kong, China
Binlong. Ye[1], [1]Department of Earth Science, The University of Hong Kong, Hong Kong, China
Josep. Ryan. Michalski[1], [1]Department of Earth Science, The University of Hong Kong, Hong Kong, China
Ashley J. King[2], [2] Planetary Materials Group, Department of Earth Sciences, Natural History Museum, Cromwell Road, London, SW7 5BD, UK

**Abstract**
The CM chondrites are characterized as primary accretionary rocks which originate from primitive water-rich asteroids formed during the early Solar System. Here, we study the mineralogy and organic characteristics of eight CM and one ungrouped chondrite to better understand their alteration history; Queen Alexandra Range 93005 (QUE 93005), Murchison, LaPaz Icefield 02333 (LAP 02333), Miller Range (MIL 13005), Mackay Glacier 05231 (MCY 05231), Northwest Africa 8534 (NWA 8534), Northwest Africa 3340 (NWA 3340), Yamato 86695 (Y-86695), and the ungrouped carbonaceous chondrite Belgica 7904 (B-7904). Raman spectroscopy has been employed to detect the presence of organic carbon in the samples, specifically through the G band at approximately 1580 cm$^{-1}$ and D band at around 1350 cm$^{-1}$. The properties of organic matter in meteorites serve as valuable indicators for characterizing the structure and crystallinity of carbonaceous materials and estimating their thermal metamorphism degree. The $R_1$ parameter, defined as the peak height ratio of the D and G bands, provides a quantifiable measure of this structural organization. Raman spectra are used to show the general mineralogy, thermal history and heating stage of CM and ungrouped chondrites. X-ray diffraction patterns further indicate the mineralogical compositions of the samples. Visible to near-infrared (VNIR) and attenuated total reflection (ATR) reflectance spectra illustrate the trends related to their mineralogy and furthermore infer aqueous alteration, thermal history of CM carbonaceous chondrites, formation, and evolution of their parent bodies.

## 1. Introduction

Carbonaceous chondrites are fragments of volatile-rich asteroids which represent the building blocks of the Solar System (Van Kooten et al. 2016; Budde et al. 2016). These meteorites are considered the most primitive objects that provide a deep knowledge about the early protoplanetary disk and the formation and evolution of the planets (Scott 2007). Since most carbonaceous chondrites were altered through aqueous alteration or thermal metamorphism on their parent bodies (Krot et al. 2005), it is vital to track the evolution of chondrite parent bodies to get insights about the initial conditions of asteroid and planet formation and their geological processes.

Aqueous alteration took place on asteroids, forming secondary minerals such as phyllosilicates, magnetite, and carbonates in the early Solar System (Brearley 2006, Suttle et al. 2021). Organic materials in meteorites are thought to have formed in interstellar space prior to the formation of the Solar System, within the solar nebula itself, and perhaps also in some asteroids (Sephton, 2004). Meteorites containing organic material probably fell to the Earth before life grew here and may indeed have been an important step to the emergence of life on Earth. Carbonaceous chondrites originating from undifferentiated objects with sizes from one to hundreds of kilometers, formed in the outer region of the main asteroid belt or beyond (Vernazza et al. 2011; Beitz et al. 2016; Tanbakouei et al. 2021). They can tell us about bodies in a specific region of the disk, and we can learn about those bodies using spectroscopy techniques. Laboratory spectroscopy techniques are increasingly used to identify the main rock-forming minerals of meteorites, and to get additional information such as physio-chemical properties of carbonaceous chondrites and primitive asteroids (Cloutis et al. 2011, 2012; Trigo-Rodriguez et al. 2014; Tanbakouei et al. 2021).

CM ("Mighei-like") chondrites are the most common and diverse group of carbonaceous chondrites (McSween Jr, 1979; Zolensky et al. 1997) and are aqueously altered rocks with typical $H_2O$ contents of ~8.7-13.9 wt.% which is structurally bound hydroxyls in phyllosilicates (Rubin et al. 2007). CM chondrites show various degrees of aqueous alteration, brecciation, and metamorphic heating, and therefore they are represented by a wide petrological range; from weakly altered (3.0) to fully altered (2.0) (Rubin et al. 2007; Hewins et al. 2014; Zolensky et al. 2020; Kimura et al. 2020; Farsang et al. 2021). CM chondrites contain up to 20 vol.% chondrules, 1–5 vol.% Calcium Aluminum Inclusions (CAIs), 4–21 vol.% mineral fragments including olivine, 1–3 vol.% Fe-sulfides and magnetite, and 57–85 vol.% phyllosilicates (Brearley and Jones 1998; Browning et al. 1998; Howard et al. 2009, 2011; Cloutis et al. 2011). Olivine compositions show a peak at $Fa_1$ (Mg-rich), but with some Fe-rich olivine (to ~$Fa_{55}$) also present (McSween 1977; Cloutis et al. 2011). An extensive interaction between liquid water and silicate rock occurred in CM parent bodies, which resulted in the hydrated mineralogy and high-water contents (Garenne et al. 2014; Suttle et al. 2021). Olivine and pyroxene are mainly present within partially altered chondrules and CAIs, and also within the matrix, where some isolated grains may have accreted directly from the protoplanetary disk (Jacquet et al. 2020; Suttle et al. 2021). The porous matrices of most CM chondrites contain clumps of minerals formed by aqueous alteration, consisting mainly of phyllosilicate and tochilinite (Trigo-Rodríguez et al. 2019). Early alteration attacked Fe-rich silicates and metal, resulting in the first phyllosilicates being Fe-rich, e.g., cronstedtite. As alteration proceeded, increasingly Fe-poor silicates were altered, producing more Mg-rich fluids and phyllosilicates (Rubin et al. 2007; Trigo-Rodríguez et al. 2019). In CM chondrites, carbonates are also widespread and typically occur as calcite, although dolomite is present in the most altered meteorites (Brearley and Jones 1998). The relative abundance of different carbonates in carbonaceous chondrites is associated with the degree of aqueous alteration (de Leuw et al. 2010).

Compared to other chondrite groups, the matrix in CM chondrites contains relatively abundant organic matter (~ 3 wt.%, Schmitt-Kopplin et al. 2010; Alexander et al. 2017). According to Kebukawa et al. (2010), the infrared spectra of organic carbon in the CM chondrites is similar to that of CI chondrites, although total carbon abundances in CMs are generally on the order of 2–4% lower than in CIs (Pearson et al., 2006). Organic compounds in CM chondrites are not found in terrestrial organisms and may have originated from extraterrestrial sources. This fact highlights the importance of studying the carbon/organic matter in CM chondrites for understanding the origins of life and the evolution of the Solar System (Pizzarello et al. 2006).

In this study, we have used Raman spectroscopy, visible/near infrared (VNIR) reflectance spectroscopy, attenuated total reflectance (ATR) and X-ray diffraction (XRD) to characterize the mineralogy and organic characteristics of eight CM and one ungrouped carbonaceous chondrite that record different degrees of aqueous and thermal alteration (Table 1). The combined datasets enable us to directly correlate meteorite mineral assemblages to spectroscopic properties, enabling interpretation of geological processes on asteroids in the early Solar System.

## 2. Samples and Experimental Methods

*2.1 Samples*

CM chondrites are generally of petrologic grade 2 due to undergoing some degree of aqueous alteration (Van Schmus and Wood 1967). A few members of the CM group have been classified as CM1 or CM1-2 as they experienced more pervasive aqueous alteration (Grady et al. 1987; Zolensky et al. 1997; Cloutis et al. 2011). CM1-2 chondrites are dominated by serpentine-group phyllosilicates in their finer-grained material and chondrules and retain few intact CAIs. According to Kimura et al. (2020) the least altered CM chondrites found to date are designated as subtype 3.0/2.9, whereas highly altered rocks which contain no mafic silicates and chondrules have been completely replaced by phyllosilicates and carbonates, are assigned to subtype 2.0 (Rubin et al. 2007). The meteorites in the present study are as follows: Queen Alexandra Range 93005 (QUE 93005), CM2.1; Murchison, CM2.5;, LaPaz Icefield 02333 (LAP 02333), CM2;  Miller Range (MIL 13005), CM1-2; Mackay Glacier 05231 (MCY 05231), CM1/2; Northwest Africa 8534 (NWA 8534), CM1/2; Northwest Africa 3340 (NWA 3340), CM2-an; Yamato 86695 (Y-86695), CM2; Belgica 7904 (B-7904), C2-ung.

Ungrouped carbonaceous chondrites, such as B-7904, are those that cannot be definitively assigned to a specific chondrite group because they have distinct chemical or mineralogical properties. Some ungrouped chondrites have experienced very little aqueous or thermal alteration, while others have experienced significant thermal and aqueous alteration which altered the organics and minerals in their matrices (Trigo-Rodríguez 2015; Tanbakouei et al. 2020). B-7904 has been studied several times, with each study proposing a close relationship to the CM group based on chemistry and mineralogy (Kojima et al. 1984; Akai 1988; Wasson et al. 1988). However, the whole-rock oxygen isotopic composition is like that of the CI chondrites (Clayton et al. 1989; Bischoff et al. 1991). B-7904 has been strongly heated after aqueous alteration and therefore contains dehydrated phyllosilicates, recrystallized secondary olivine and melted Fe-sulfides (King et al. 2019). B-7904 consists of partially altered chondrules, hydrous and anhydrous minerals in the matrix and phyllosilicate clasts, with a variety of olivine-rich chondrules and aggregates (Kimura and Ikeda 1992; Ikeda et al. 1993) which experienced peak metamorphic temperatures > 750 °C (Tonui et al. 2014). Y-86695 also experienced post-hydration thermal metamorphism. According to Nakamura (2005; 2006), Y-86695 is a Stage II (< 500°C) sample (Nakamura 2005; 2006), although reflectance spectra of Y-86695 show absorption bands near 0.85, 1.05, and 1.25 µm characteristic of olivine (Cloutis et al. 2012). This could reflect a relatively low degree of aqueous alteration (King et al. 2021) and/or thermal metamorphism and recrystallisation of secondary olivine at temperatures >500°C.

A wide range of CM carbonaceous chondrites were selected because they have experienced aqueous alteration and contain some primitive organic matter of interest to the current study. Nine samples have been selected from the collection of Antarctic Search for Meteorites (ANSMET), the National Institute of Polar Research (NIPR) and Arizona State University (ASU) (Table 1). The same powders (except for MIL 13005 and NWA 3340) were previously studied using XRD, thermogravimetric analysis (TGA), and IR spectroscopy by King et al. (2017, 2019, 2021) and Bates et al. (2020). For this study, around 100 to 300 mg of eight of the powders was shaped into a pellet to acquire a mineral map. Although Raman scattering can be observed using a powder sample, a pellet sample can provide Raman spectra with a better signal-to-noise (S/N) ratio. Powder was pressed into pellets using a die and hydraulic oil press with the maximum load of 5 tons and the load on the die was held for 50 seconds to 1 minute. For MIL 13005 we instead received a small chip on which we performed Raman mapping rather than producing a powdered pellet.

| No. | Sample | Petrologic Type | Sample Preparation | Exposure time (s) | Laser Power (mW) | Step (μm) | Area (μm × μm) |
|---|---|---|---|---|---|---|---|
| 1 | MIL 13005 | CM1-2 | Chip | 5 | 2.3 | 4 | 46×188 |
| 2 | NWA 8534 | CM1/2 | Pellet | 10 | 2.3 | 8 | 784×480 |
| 3 | LAP 02333 | CM2 | Pellet | 10 | 2.3 | 5 | 650×375 |
| 4 | MCY 05231 | CM1/2 | Pellet | 10 | 4.6 | 8 | 800×440 |
| 5 | QUE 93005 | CM2.1 | Pellet | 10 | 4.6 | 8 | 800×472 |
| 6 | Y-86695 | CM2 | Pellet | 10 | 4.6 | 9 | 819×495 |
| 7 | Murchison | CM2.5 | Pellet | 10 | 4.6 | 9 | 243×279 |
| 8 | NWA 3340 | CM2-an | Pellet | 10 | 2.3 | 10 | 820×520 |
| 9 | B-7904 | C2-ung | Pellet | 10 | 4.6 | 8 | 824×528 |

Table 1. Raman mapping measurement parameters. The excitation laser for all of them was 532 nm.

2.2 Raman spectroscopy

Raman spectroscopy has been used as a powerful tool for mineral identification, assessing aqueous alteration, degree of shock and different forms of carbon in carbonaceous chondrites (Haskin et al., 1997; Beyssac et al. 2004; Litasov et al. 2017; Potiszil et al. 2021). It is also a sensitive indicator of water and can also identify the minerals in which $OH/H_2O$ are hosted. Raman spectroscopy also provides information about the rock texture and can identify alteration products of primary minerals (Yui et al. 1996; Haskin et al. 1997; Brolly et al. 2016; Potiszil et al. 2021).

In Raman spectroscopy, a monochromatic light from a source is scattered inelastically from the minerals in the target. The spectrometer rejects reflected light from the source and Rayleigh-scattered light of that same wavelength, and it analyzes the longer wavelengths of the Raman-scattered light (Stokes lines, $\lambda + \Delta\lambda$). The difference in wavelength between the source light and the scattered light corresponds to transition energies in the material that produced the scattering. This difference in wavelength, $\Delta\lambda$, normally given in units of wave number $cm^{-1}$, is called the Raman shift.

To study the different mineralogical phases and to provide possible information regarding the thermal history of carbonaceous chondrites, micro-Raman spectra with a spot size of ~0.86 μm and laser power of 4.6 mW (10% laser power) or 2.3 (5% laser power) were obtained to ensure that laser induced heating did not alter the meteoritic organic matter. The 2D maps were acquired as grids with 5-10 μm spacing, each point consisting of 1 acquisition recorded over 4-10 s. The Raman measurements were carried out at room temperature using an inVia Raman imaging spectrometer (Renishaw Company) that has high spectral (1 $cm^{-1}$) resolution. Raman spectra were collected using an excitation wavelength of 532 nm.

The Raman spectrometer operated between 200 and 1900 $cm^{-1}$. The appearance of distinct peaks in this spectral region allows us to distinguish many of the major rock-forming minerals. The precise location of the peaks in a Raman spectrum is sensitive to the chemical bonds involved, and is thus sensitive to various factors, including the effects of shock on the lattice structure. To calibrate our Raman spectrometer, we used the sharp peak of stress-free Si at 520.3 $cm^{-1}$ (Sik Yoo et al. 2014). It should be emphasized that preparation of pellets might potentially modify the sample being analyzed. The process can cause physical changes to the sample, such as changing its crystalline structure, altering particle size and shape, and introducing strain and defects. The Raman scattering process is highly sensitive to the local environment and structure of the sample, and changes to the sample's physical or chemical properties can affect the Raman spectra. Similarly, pressing a sample into a pellet can cause densification, stress, or deformation, which can modify the Raman spectra (McCreery, 2005; Kneipp et al. 2006). However, the overall mineralogy and composition of the sample shouldn't be significantly affected.

The pellets were scanned with a 50x long working distance objective, and different time exposures per spectrum were used (Table 1). The baseline of the spectra for each sample was subtracted, and the D and G band properties were determined using a curve-fitting procedure. A curve-fitting technique was employed to extract quantitative information from the Raman spectra. This technique involves fitting a mathematical model, such as a Gaussian or Lorentzian function, to the experimental data to determine the underlying components that contribute to the spectrum. The choice of the appropriate model was based on the goodness of fit, as assessed by the reduced chi-square value and other statistical measures. Prior to curve-fitting, the spectra were pre-processed, which involved steps such as smoothing, normalization, and background subtraction to enhance the signal-to-noise (S/N) ratio and improve the accuracy of the fitting results. Factors such as the S/N ratio, the number of peaks, and the shape of the peaks can all impact the accuracy of the curve-fitting results. In this work, curve-fitting was applied to Raman peaks using the Renishaw WIRE software, with user-defined initial peak positions, widths, and constraints to guide the fitting process. The resulting fitted parameters, such as peak positions, intensities, and widths, were then used to analyse and compare the D and G band properties of the samples.

There are no characteristic features of phyllosilicates in Raman spectra of meteorites in the spectral range analysed; they typically have features in the range of 3600-3800 $cm^{-1}$ (Wang et al. 2015). However, phyllosilicates were detected by XRD, VNIR and ATR spectroscopy in this study.

2.3 X-ray diffraction

X-ray diffraction (XRD) patterns were collected for the qualitative identification of minerals present in each sample. For each sample (except LAP 02333), a small amount (several tens of milligrams) of powder was placed on a zero-background silicon substrate for analysis. The XRD pattern of LAP 02333 was instead collected using a pellet made from its fine powders. Co K$\alpha$1 ($\lambda$= 1.7890Å) radiation was employed to avoid the high background due to Fe atoms fluorescing in the X-rays produced by Cu K$\alpha$1 radiation. All measurements were conducted on a Rigaku MiniFlex Benchtop X-ray diffractometer operated at 40 kV and 15 mA for scanning between 3° and 80° (2$\theta$) with a scanning rate of 2° per minute and a step of 0.02° at the Planetary Spectroscopy and Mineralogy Laboratory (PSML) at the department of Earth Science in the Science Faculty at University of Hong Kong. Each measurement took about 40 minutes. X-ray diffractograms were compared with the PDF-4/Minerals database in the JADE Pro software (ICDD) to identify the mineralogical compositions. Semi-quantitative analyses were conducted using the whole pattern fitting (WPF) function provided in JADE Pro.

2.4 Visible to Near IR Spectroscopy

Visible/near-infrared (VNIR) reflectance spectra of fine meteorite powders (grainsize typically <50 μm) were collected at the PSML using a Nicolet iS50 interferometer with an Indium Gallium Arsenide (InGaAs) detector and Si detector under a continuous flow of 99.999% nitrogen gas to minimize the interference of water vapor. A diffuse reflectance accessory, Praying Mantis, was attached to the Nicolet iS50 interferometer to collect spectra from 400 to 2500 nm with a spot size of ~ 3 $mm^2$ and integration of 300 scans. Spectralon in the visible region from Labsphere Inc. and pure KBr powder in the NIR region were used as the background for measurements in the visible and NIR regions, respectively. Approximately 50 mg of fine meteorite powders were drawn over the surface of the Spectralon to generate an almost flat surface without pressing the uppermost layers of the samples. A sharp peak in the spectra around 633 nm wavelength is caused by the HeNe laser of the interferometer and data from 630-635 nm were averaged in later data processing. The spectral range is comparable to that collected from space- and ground-based telescopes and deep space exploration missions (e.g., Hayabusa2 and OSIRIS-REx) targeting asteroids. To determine the band depths (absorption strengths) of the relevant absorptions, continuum-removed spectra were analyzed using the techniques outlined by Clark and Roush (1984). A third-order polynomial was utilized to identify the local absorption minimum at the centre of the feature. The estimated uncertainties are one standard deviation due to variability from the average of calculated parameters (Bates et al., 2020).

2.5 ATR Spectroscopy

Attenuated Total Reflectance (ATR) mid-infrared spectra were collected on the same samples for which VNIR spectra were collected, from 400 to 4,000 cm$^{-1}$ on a PerkinElmer Spectrum Two FTIR Spectrometer at the Department of Chemistry of The University of Hong Kong using 4 cm$^{-1}$ spectral sampling and 32 scans integration with deuterated triglycine sulphate (DTGS) detector with a KBr beam splitter.

## 3. Results & Discussion

3.1. General mineralogy of CMs determined by XRD

The main crystalline phases of each sample were identified based on powder XRD patterns (Fig. 1 and Fig. 2). In the CM1/2 group, the meteorite MCY 05231 contains olivine, pyroxene and serpentines, in which Fe-rich serpentines account for two-thirds. Other minor phases include magnetite, iron- and nickel-sulfides, gypsum and calcite. MIL 13005 is dominated by magnetite, iron- and nickel-sulfides and a low percentage of serpentines. Olivine, pyroxene and gypsum are also present. NWA 8534 is mainly composed of olivine, pyroxene and Mg-rich serpentine with a little iron-sulfide.

In the CM2 group, the meteorite LAP 02333 contains olivine, pyroxene and serpentines, in which Mg-rich serpentines are approximately ten percent more abundant than Fe-rich serpentines. Other minor phases include magnetite, iron- and nickel-sulfides, gypsum and calcite. The meteorite Murchison is mainly composed of olivine, pyroxene, Mg-rich serpentine and iron-sulfides with a little magnetite. The meteorite QUE 93005 is dominated by Mg-rich serpentines and magnetite. Other minor phases include olivine, iron- and nickel-sulfides and gypsum. No pyroxene was identified in this sample. The meteorite Y-86695 consists of olivine, pyroxene, magnetite and iron-sulfides with minor Mg-sulfates and calcite. No crystalline phyllosilicates were identified in this sample. The meteorite NWA 3340 is dominated by Mg-rich serpentines. Other phases include olivine, pyroxene, iron- and nickel-sulfides and anhydrite. The meteorite B-7904 is the only sample belonging to the group C2-ung and it is primarily composed of olivine. Other phases include pyroxene, iron- and nickel-sulfides, gypsum and a small amount of Fe-Ni metal. The overall mineralogy of the analyzed samples is in good agreement with the previous XRD studies on the same powders conducted by King et al. (2017; 2021). XRD patterns for MIL 13005 and NWA 3340 have not been previously reported in the literature. XRD patterns of MIL 13005 and Y-86695 are consistent with them being stage II (300-500C) samples (i.e., they contain poorly crystalline, dehydrated phyllosilicates), and B-7904 is a stage IV (>750C).

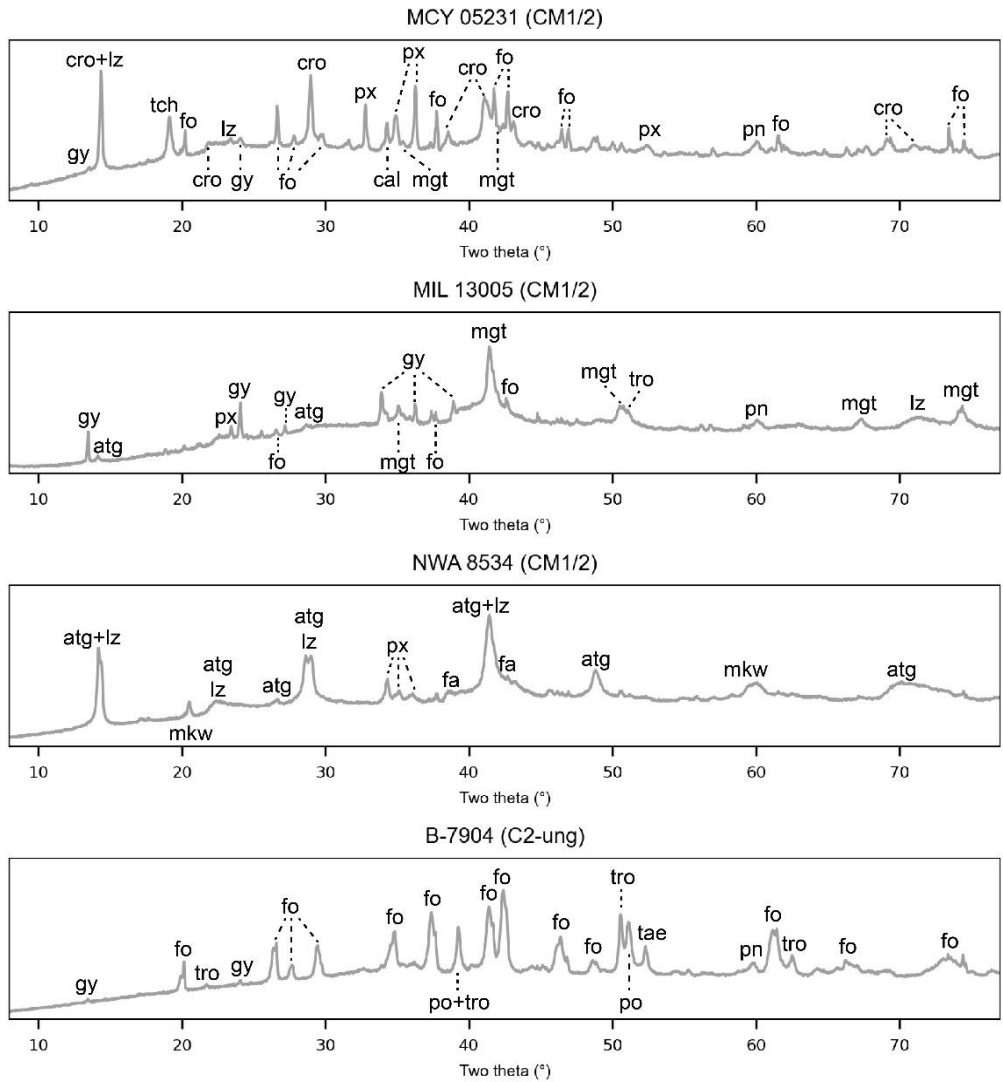

Figure 1. The XRD patterns of CMs in the CM1/2 and C2-ung groups.
Olivine: fo – forsterite, fa – fayalite; pyroxene: px – pyroxene (including enstatite, clinoenstatite, diopside, pyroxene-ideal); serpentine: lz – lizardite, atg – antigorite, cro – cronstedtite; iron oxides: mgt – magnetite; sulphides: tro – troilite, po – pyrrhotite, tch – tochilinite, mck – mackinawite, pn – pentlandite; sulfates: gy – gypsum; carbonates: cal – calcite; metal: tae – taenite.

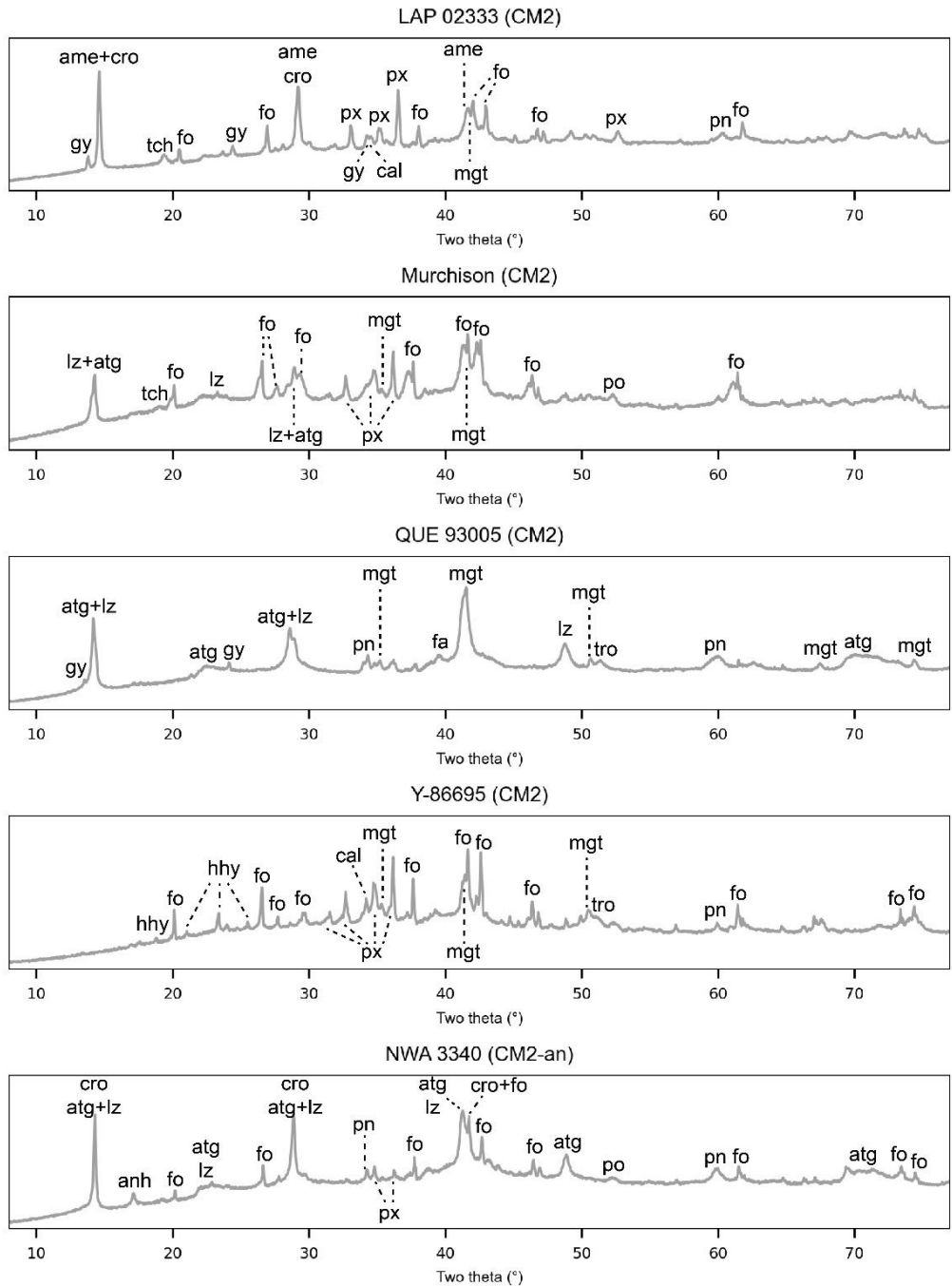

Figure 2. The XRD patterns of CMs in the CM2 and CM2-an groups.
Olivine: fo – forsterite, fa – fayalite; pyroxene: px – pyroxene (including enstatite, clinoenstatite, diopside); serpentine: lz – lizardite, ame – amesite; atg – antigorite, cro – cronstedtite; iron oxides: mgt – magnetite; sulphides: tro – troilite, po – pyrrhotite, tch – tochilinite, mck – mackinawite, pn – pentlandite; sulfates: gy – gypsum, hhy - hexahydrite; carbonates: cal – calcite.

## 3.2. Organic matter and mineralogy inferred from spectroscopic techniques

### 3.2.1. Raman spectra

Raman spectroscopy has previously been used to study the mineralogy and alteration processes in carbonaceous chondrites (Quirico et al. 2018; Nascimento-Dias et al. 2021; Potiszil et al. 2021). In this study, Raman spectra of most carbonaceous chondrites, either as chips or pellets, consist of broad carbon bands, the so-called D and G bands. Moreover, evaluation of the D and G band peak centers and full width half-maximum (FWHM) values (Table 2), can reveal the thermal history of meteorites and their parent bodies, and also the nature of bonding within carbonaceous materials such as meteoritic organic matter. According to Busemann et al. (2007), the presence of highly disordered carbon in more primitive meteorites, including CI, CM and CR carbonaceous chondrites, is the result of UV or particle irradiation that occurred before the organic matter was incorporated into the parent body of these meteorites. Precursor composition in the chondrites containing more primitive organic matter has a significant impact on the Raman response of organic matter (Quirico et al. 2009; Quirico et al. 2018; Potiszil et al. 2021).

Raman mapping of a chip of the carbonaceous chondrite MIL 13005 is provided in Figure 3..Based on the presence of the D- and G-bands in the Raman spectra, carbonaceous matter appears to be closely associated with the minerals, considering that only certain minerals can be detected from the selected spectral ranges. The Raman peak at 983 cm$^{-1}$ is attributed to hydrated phosphate, which might belong to a distinct structural type, depending on its hydration content and the atomic weight of the phosphate (Clavier et al. 2018). Phosphates in hydrous carbonaceous chondrites are thought to be secondary minerals, formed during parent body hydrothermal processing (Flynn et al. 2022). Phosphates show variable morphologies, and they often occur as aggregates, or in the CI chondrites are associated with pyrrhotite, magnetite, and carbonate (Alfing et al. 2019), as appears to be the case here. The basic characteristics of phosphate minerals include the strong $v_1$ symmetric stretching mode of PO$_4$ at 950–990 cm$^{-1}$, and here the assigned wavenumber is at 983 cm$^{-1}$ (Litasov et al. 2017; Clavier et al. 2018).

In some restricted fragments with high Ca-concentration, alteration of CAIs might form abundant carbonates or phosphates. Ca and P could also be leached from chondrule glass, and P is present in metal, which is one of the first phases to be altered. Phosphates are observed surrounding CAIs in metamorphosed chondrites, however there's no evidence that MIL 13005 experienced high temperatures (Ward et al. 2017; Ondrejka et al. 2018). XRD indicates that MIL 13005 experienced short-lived heating at 300 – 500 °C.

Representative Raman spectra of other analyzed samples are presented in Figures 4 and 5. The Raman spectrum of calcite in QUE 93005 includes one broad peak at 289 cm$^{-1}$ and one sharp peak at 1094 cm$^{-1}$. Peaks at ~213 and ~270 cm$^{-1}$ are assigned to Iron III hydroxide, which is found in B-7904, LAP 02333, MCY 05231, NWA 3340, QUE 93005 and Y-86695. The Raman spectra of all eight CM samples and one ungrouped sample show peaks from forsterite at ~822 and ~857 cm$^{-1}$. Magnetite shows a broad peak around ~533 cm$^{-1}$ or 670 cm$^{-1}$ and is present in B-7904, LAP 02333, MCY 05231, NWA 3340 and MIL 13005. Diopside shows specific peaks at 231, 321, 670 and 1008 cm$^{-1}$ in Raman spectra of LAP 02333. The spectra of all these minerals also contain broad, low-intensity D and G bands at ~1350 and ~1590 cm$^{-1}$, respectively, corresponding to organic carbon.

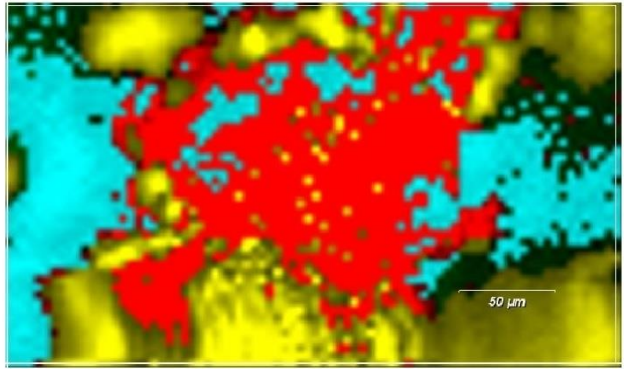
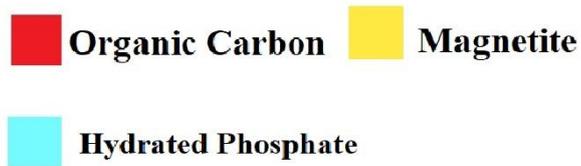
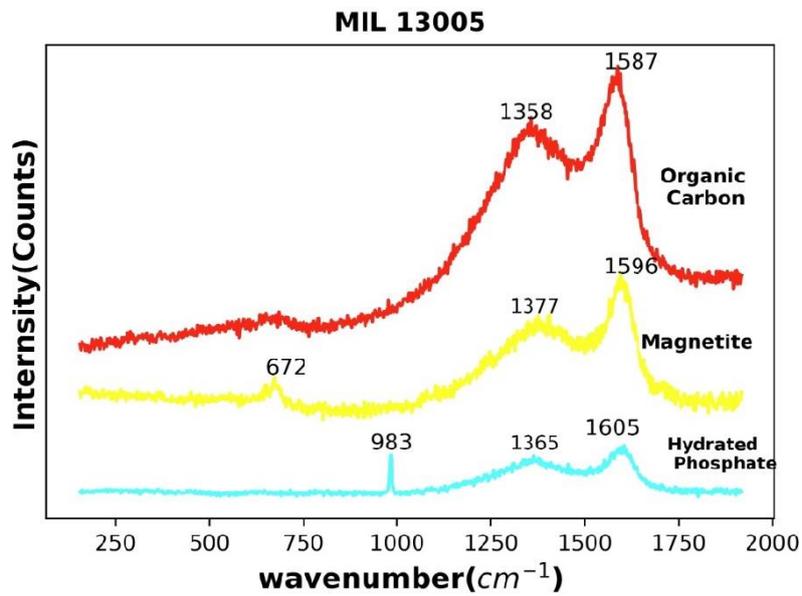

Figure 3. Compositional maps and Raman spectra illustrating the various minerals in a 8648 (46×188) µm² sized area on a chip of the CM chondrite MIL 13005. Note that phyllosilicates do not have characteristic Raman features in the spectral range analysed.

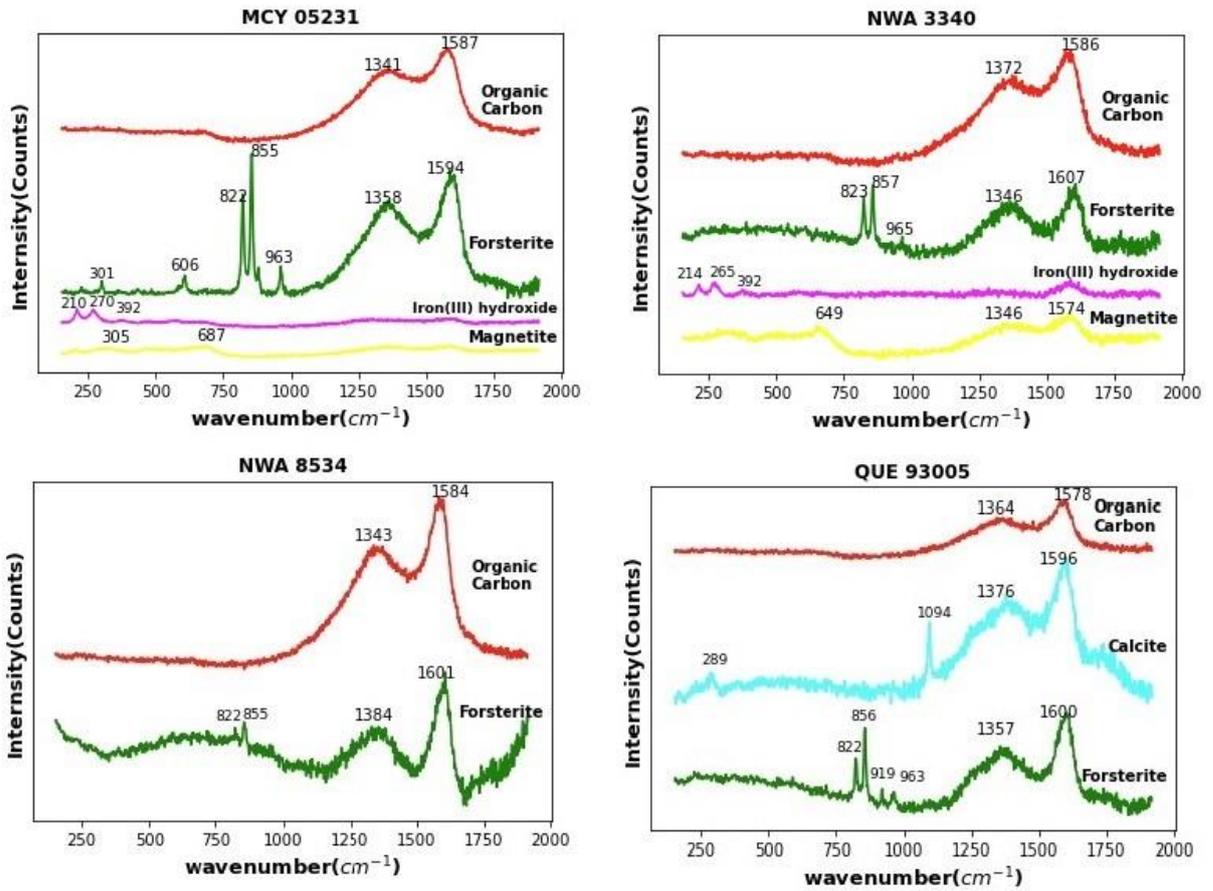

Figure 4. Raman spectra of different minerals found in meteorite samples: MCY 05231, NWA 3340, and QUE 93005. The samples were prepared as pellets.

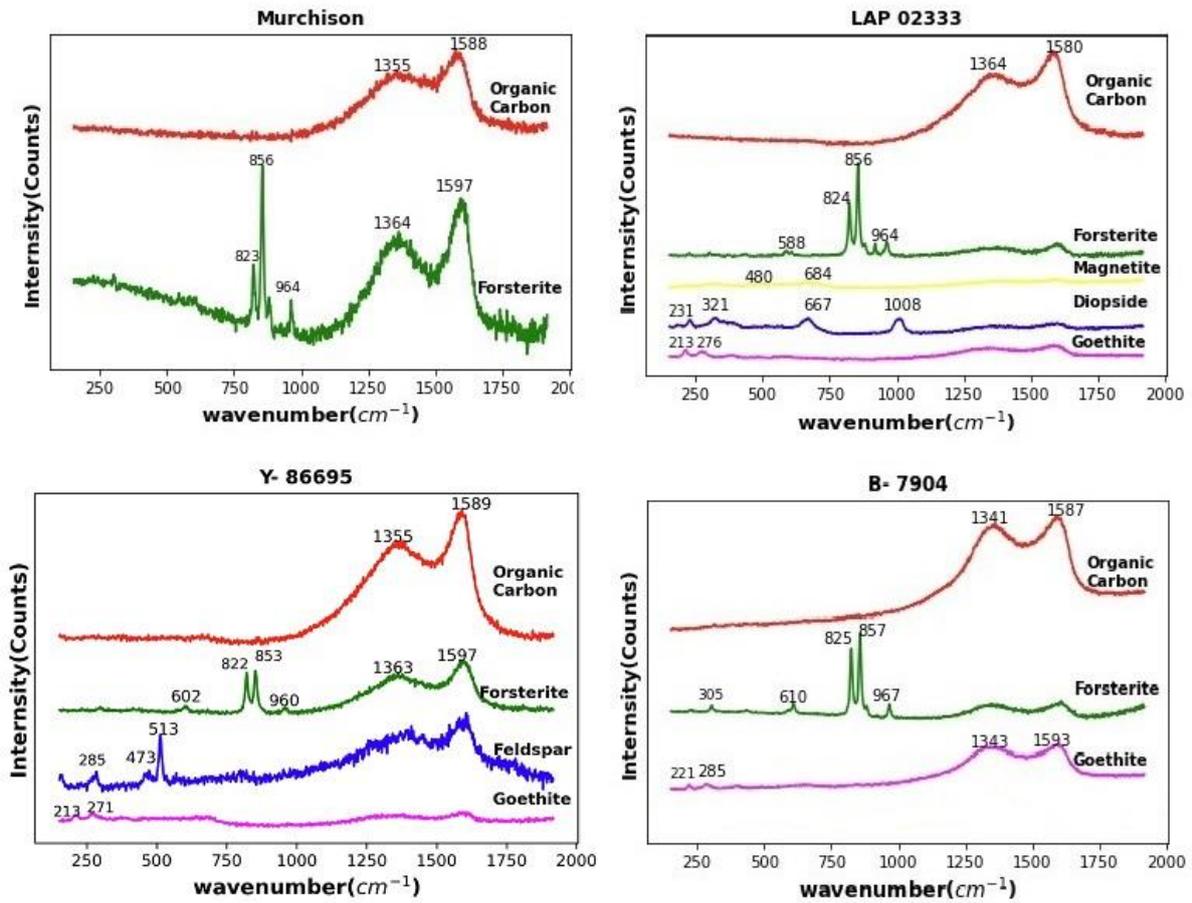

Figure 5. Raman spectra of different minerals found in meteorite samples: Murchison, LAP 02333, Y-86695 and B-7904. The samples were prepared as pellets. The Raman spectra of mineralogy found in B-7904 as an ungrouped carbonaceous chondrite resembles the Raman spectra of minerals in CMs.

*3.2.2. Thermal metamorphism suggested by carbonaceous materials in CMs*

Carbon is common within extraterrestrial materials, appearing as both soluble and insoluble organic matter in primitive meteorites (Busemann et al. 2006). In meteorites, carbon can be hosted in the matrix or sometimes, a small amount of carbon is present within metal grains (Fe-Ni) (Dubessy J et al. 2012). Carbonates are ubiquitous in CMs and represent a minor proportion of secondary mineral assemblages (up to 2–3 vol.%; Lee et al. 2014; Vacher et al. 2018). Various processes such as thermal metamorphism on the parent body or terrestrial weathering may affect the structure of the insoluble organic carbon (Busemann et al. 2009; Dubessy J et al. 2012).

According to Suttle et al. (2021), one third of CM chondrites experienced post aqueous alteration thermal metamorphism, with impacts likely to be the main source of heating. As metamorphism causes the escape of volatiles, metamorphosed CMs have a dehydrated mineralogy, graphitized organic matter, and depleted $H_2O$, C and N contents. The CM parent bodies were highly shocked by collisions, which is shown by the dominance of CM micrometeorites among the extraterrestrial dust flux and the presence of CM-like rubble pile asteroids (e.g., Bennu and Ryugu). Bennu shares significant similarities with CM chondrites, as both exhibit widespread hydrated minerals, such as phyllosilicates, and the presence of a 2.7 µm absorption band, which is a characteristic feature of hydrated minerals. The reflectance spectra of LAP 02333 and Bennu has similar infrared spectral features in the range of MIR region as Bennu showing broad Si-O stretching absorption with weak Mg-OH bending around 600 $cm^{-1}$ region (Hamilton et al. 2019; Amsellem et al. 2020). The overall brightness and shape of VNIR spectra of Ryugu's surface are similar to thermally metamorphosed CI samples and shocked CMs (Kitazato et al. 2019), however no evidence for significant heating has been reported for the Ryugu samples (Naraoka et al. 2023).

The peak parameters of the D and G bands, such as the peak center locations, peak widths (FWHM), and peak intensity ratios can be used to estimate the structure and crystallinity of carbonaceous materials and the degree of thermal metamorphism (Busemann et al. 2007; Chan et al. 2019). If the organic matter is perfectly crystalline, the G band is the only peak which arises from stretching vibrations found within layers of graphite-like material and is present at ~1580 $cm^{-1}$ (Tuinstra and Koenig 1970). Defects in the crystal lattice of aromatic organic matter are represented by the D band at ~1350 $cm^{-1}$ (Ferrari and Robertson 2000; Busemann et al. 2007; Potiszil et al. 2021). The presence of the D band indicates the presence of disordered carbon.
The structural organization of organic carbon can be quantified through the $R_1$ and $R_2$ parameters, defined as the peak height ratio of the D and G bands and FWHM ratio of D and G band, respectively (Table 2). The numerical values of the derived spectral parameters of the D and G bands, including their FWHM, peak positions and intensity ratio ($R_1$) are presented in Fig. 6. These suitable parameters, which are temperature dependent, could identify trends in Raman spectra (Homma et al. 2015; Visser et al. 2018). In Fig. 6a, c, d and e it is clearly shown that B-7904 is an exception heated at >750 °C. Lower FWHM-D values are associated with increased heating, as observed for B-7904 in all plots. Additionally, higher R1 values also indicate the presence of heating. Other samples are similar, which is consistent with the Quirico et al. (2018) and Kiryu et al. (2020) studies. Although LAP 02333 is an outlier, there is no strong evidence of heating from XRD (e.g., serpentine and tochilinite reflections) or other Raman parameters for this chondrite. The graph in Fig 6b can't easily resolve between different heating stages, but it is consistent with Quirico et al. (2018). According to Kiryu et al. (2020), the FWHM-D of chondrites decreased with increasing temperature of the parent body (Fig. 6).

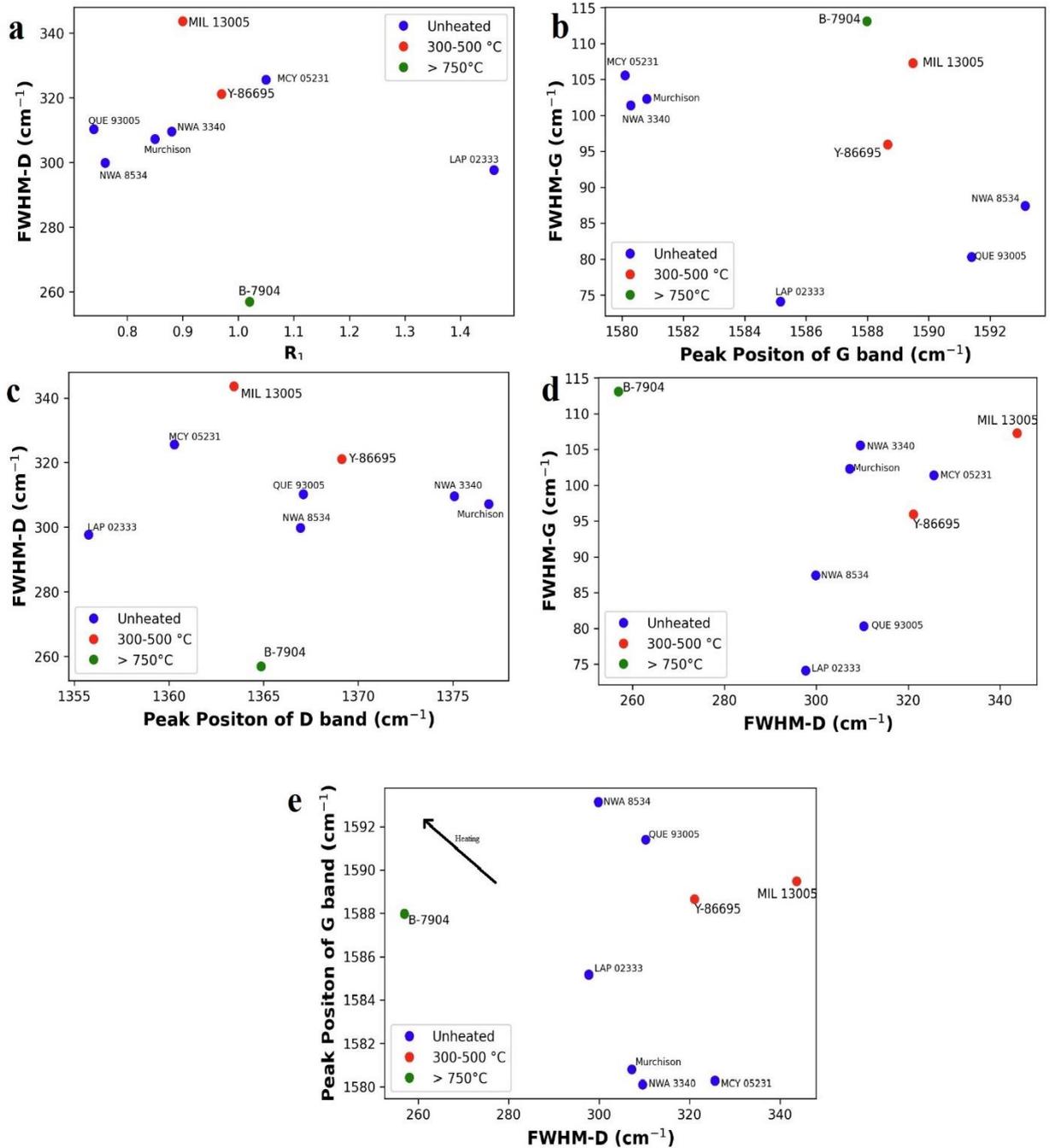

Figure 6. Spectral parameters of studied chondrites determined from Raman spectra: a) FWHM-D vs. $R_1$ (the peak intensity ratio of the D and G bands), B-7904 is clearly an outlier, heated at >750 °C. The others are similar, which is consistent with the Quirico et al. (2018) and Kiryu et al. (2020) studies. b) FWHM-G vs. peak position of G band, can't easily resolve between different heating stages, but it is consistent with Quirico et al. (2018). c) FWHM-D vs. peak position of D band. Here also B-7904 is clearly heated at >750 °C. The others are similar consistent with the findings of Quirico et al. (2018) and Kiryu et al. (2020). d) FWHM-G vs. FWHM-D. B-7904 is heated >750 °C. e) peak position of G band vs. FWHM-D, B-7904 is an outlier, heated at >750 °C. The others are consistent with the Quirico et al. (2018) and Kiryu et al. (2020) studies.

Regarding the thermal history of samples, Raman seems to be more sensitive to low levels (<300 °C) of thermal metamorphism (King et al. 2021) that cannot be detected from XRD analysis. Highly heated

CMs show structural modifications that result in narrower D-band widths, with B-7904 the only sample in this study where the FWHM-D-width is an outlier consistent with strong thermal metamorphism (Fig. 6) (Tonui et al. 2014; King et al. 2019). XRD confirms the fact that B-7904 was strongly heated as the experiment was applied on powders with no modification. As B-7904 experienced post-hydration thermal metamorphism, it contains few/no phyllosilicates. Dehydrated mineralogy suggests that heating occurred in the meteorites parent bodies which indicates that hydrated asteroids have been heated to a certain degree (Nakamura T. 2005). The XRD results here and in previous work also confirm this.

According to Busemann et al. (2007), highly metamorphosed chondrites, like Y-86695 (CM2) a stage II (300 – 500 °C), and B-7904 (C2-ung) a stage IV (>750 °C) heated meteorite, have higher peak intensity value of D band (Table 2; Busemann et al. 2007; Kiryu et al. 2020). The G band position values increased as thermal metamorphism developed (Busemann et al. 2007; Chan et al. 2019). NWA 3340 and LAP 02333 have lower G band values than MIL 13005, Y-86695 and B-7904. Although they were subjected to heating at lower temperatures, it is likely that long-term heating occurred in their parent bodies (Busemann et al. 2007).

| No. | Sample | Peak Center D band | Peak Center G band | FWHM D band | FWHM G band | $R_1$ | $R_2$ |
|---|---|---|---|---|---|---|---|
| 1 | MIL 13005 | 1363.4± 4.1 | 1589.5± 5.5 | 343.7± 9.0 | 107.3± 2.8 | 0.90 | 3.20 |
| 2 | NWA 8534 | 1367.0± 5.3 | 1593.1± 5.1 | 299.9± 6.4 | 87.4± 3.5 | 0.76 | 3.43 |
| 3 | LAP 02333 | 1355.8± 3.0 | 1585.2±3.5 | 297.7± 7.2 | 74.1± 6.6 | 1.46 | 4.01 |
| 4 | MCY 05231 | 1360.3± 3.2 | 1580.3± 2.6 | 325.6± 6.3 | 101.4± 8.3 | 1.05 | 3.20 |
| 5 | QUE 93005 | 1367.1± 3.5 | 1591.4± 2.5 | 310.3± 8.6 | 80.3± 5.6 | 0.74 | 3.86 |
| 6 | Y-86695 | 1369.1± 5.1 | 1588.7± 4.0 | 321.1± 8.0 | 96.0± 4.8 | 0.97 | 3.34 |
| 7 | Murchison | 1376.9± 2.7 | 1580.8± 3.6 | 307.2± 9.2 | 102.3± 5.3 | 0.85 | 3.00 |
| 8 | NWA 3340 | 1375.1± 3.0 | 1580.1± 4.5 | 309.6± 9.8 | 105.6± 5.4 | 0.88 | 2.93 |
| 9 | B-7904 | 1364.9± 4.3 | 1588.0± 4.1 | 256.9± 7.3 | 113.1± 3.2 | 1.02 | 2.27 |

Table 2. The table presents the numerical values for the peak centers and FWHM of the Ramn shift D and G bands (/cm$^{-1}$) assigned to each meteorite sample. Each value is an average of 10 randomly selected points. $R_1$ is peak intensity ratio of the D and G bands, $R_2$ is peak width (FWHM) ratio of the D and G band. The uncertainties on the Raman spectral parameters are 1σ standard deviation.

### 3.3. Evidence of Phyllosilicates in CMs

*3.3.1. Phyllosilicates and organic matter determined by ATR spectroscopy*

The ATR mid-infrared spectra of the CM samples are shown in Figure 7. Due to samples being analysed at ambient air environment and room temperature, there is a broad feature from 3800 – 3000 cm$^{-1}$ caused by the adsorbed atmospheric water molecules (Beck et al. 2010). The main features are absorption at 1000 - 700 cm$^{-1}$ and minor absorptions around 600 cm$^{-1}$ and 400 cm$^{-1}$, providing important information on the nature of silicate phases in the samples. The 1000-700 cm$^{-1}$ feature in CM meteorite spectra is interpreted to be the Si-O stretching, while the 600-400 cm$^{-1}$ wavelength region is associated with structural changes in phyllosilicates, which are caused by the bending vibration for tetrahedral SiO$_4$ groups as well as lattice deformation modes (Michalski et al. 2005). Fig. 8 shows the NIR spectra of CM meteorites, which clearly indicate anhydrous silicates (olivine) and hydrated phyllosilicates. However, we notice that phyllosilicates in CM chondrites are complicated and different from our terrestrial endmember standards, which is most likely due to different formation conditions (temperature and pressures) or protoliths (e.g., Calvin and King 1997; Howard et al. 2009). The Reststrahlen band of CM1/2 and CM2 range from 993 to 941 cm$^{-1}$ and their overall spectral shape is similar to Fe/Mg serpentine with a varied amount of olivine, which indicates the different degree of alteration. In general, less altered meteorites typically have a higher abundance of olivine present, which causes a dominate absorption at lower wavenumber for Si-O stretching (McAdam et al. 2015; Bates et al. 2020, 2021; King et al. 2021). Highly altered meteorites contain higher abundances of phyllosilicates, and their Si-O stretching feature is therefore dominated by a longer wavenumber absorption near 875 cm$^{-1}$. MIL 13005 exhibits absorption around 980 cm-1, which may be attributed to a structural change caused by the dehydration of phyllosilicates, indicating heating. Moreover, the presence of additional mineral phases could shift this absorption to a longer wavenumber region, likely due to sulfate, as inferred from the XRD results. Sulphate generally shows strong absorption features around 1120-1150 cm$^{-1}$ and there is a shoulder around 1120 cm$^{-1}$ for MIL 13005. This sample also shows a weak absorption around 560 cm$^{-1}$. It indicates the presence of magnetite and is consistent with a significant portion of magnetite suggested by the XRD pattern and Raman mapping. Magnetite abundances are generally higher in the most aqueously altered carbonaceous chondrites (King et al. 2015; 2017). MIL 13005 also shows a prominent absorption feature at 952 cm$^{-1}$ and strong 0.7 and 2.3 μm absorption features. The presence of magnetite affects the NIR and shortwave infrared (SWIR) wavelength region and it's difficult to evaluate the degree of alteration using only a single spectral feature. NWA 8534 has an absorption feature located at 948 cm$^{-1}$, QUE 93005 at 942 cm$^{-1}$, and MCY 05231 at 932 cm$^{-1}$. Murchison presents a doublet absorption feature at 952 and 880 cm$^{-1}$ while LAP 02333 shows a broad and weak absorption feature around 921 cm$^{-1}$. The LAP 02333 ATR spectrum is quite different from other meteorites, which might be related to its unusual Raman properties shown in Fig. 6. B-7904 experienced strong thermal metamorphism as suggested by XRD and Raman results and it shows an unusual shape in the mid-IR wavelength region. The strong absorption feature of B-7904 shifts to 830 cm$^{-1}$ and 860 cm$^{-1}$ with a shoulder around 482 cm$^{-1}$, which is consistent with abundant olivine formed through thermal metamorphism (Nakamura 2005; King et al. 2019). Overall, the most altered samples generally exhibit a single smooth U-shape-like spectra at the Si-O stretching region. In addition, weak absorption located at 2850 and 2925 cm$^{-1}$ caused by CH$_2$ symmetric and asymmetric stretching bending of aliphatic organic matter are seen in MIL 13005, Murchison, QUE 93005, and NWA 3340, further supporting the organic matter detected by Raman considering ATR is not as sensitive to organics as Raman spectroscopy.

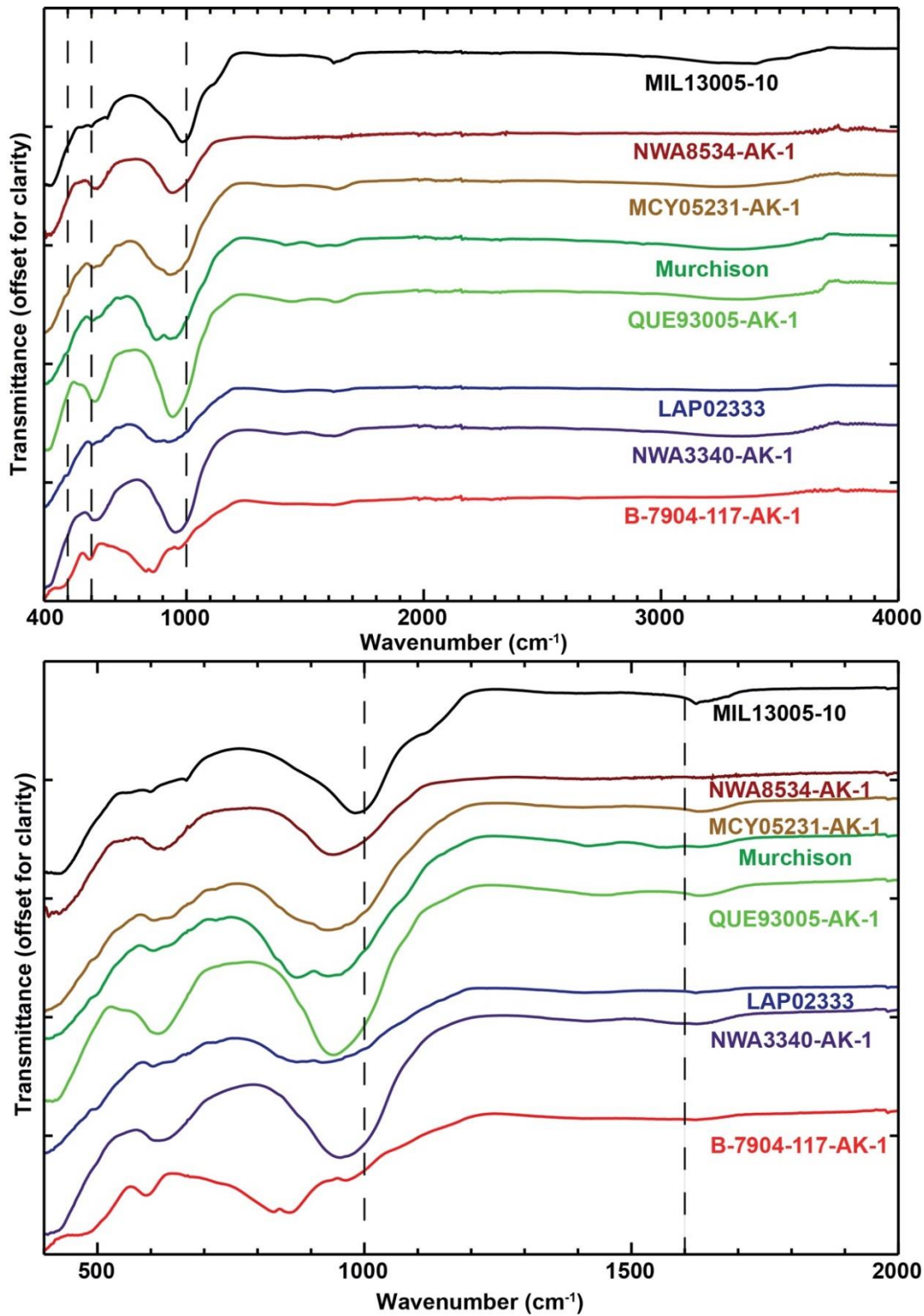

Figure 7. The upper panel shows the Attenuated Total Reflectance (ATR) spectra of CM meteorite samples, and the lower panels shows the enlarged view of the wavelength region from 400 to 2000 cm$^{-1}$.

*3.2.2 Phyllosilicates and other minerals determined by VIS to NIR spectroscopy*

Spectral reflectance properties from 0.4 to 2.5 μm of CM carbonaceous chondrites are presented in Figs. 8 and 9. The VNIR spectra of CM meteorites are generally dark, often exhibiting strong absorption bands in the 0.7 μm region caused by $Fe^{3+}$-$Fe^{2+}$ charge transfer, suggesting the presence of phyllosilicate minerals, specifically serpentine-group as previously reported by other studies (e.g., Cloutis et al. 2011; Beck et al. 2018; McAdam et al. 2015). The band depth and band minimum of the 0.7 µm feature is variable, which might suggest that the strength and position of the charge transfer band may represent a complicated mixture of ferric to ferrous iron ratio in both Mg- and Fe-rich phyllosilicates, Mg-bearing phyllosilicates and Fe hydroxides. The spectra of MIL 13005 and Y-86695 in this region differ from the other samples (Figure 8), which is consistent with their XRD patterns not showing strong peaks from hydrated phyllosilicates, likely because they experienced later thermal metamorphism (Nakamura 2005; King et al. 2021). However, for MIL 13005, the attenuated 0.7 µm absorption feature may also be explained by a high magnetite abundance, as indicated by its XRD pattern (Fig. 1).

In the SWIR region, NIR redslope are observed and absorption bands are difficult to observe, although the continuum removed spectra highlight a few features (Figure 9). CM1/2, CM2, but not the C2-ung meteorites, exhibit varied absorption features at 1.4, 1.9, and 2.3 µm, which correspond to $H_2O$ stretching overtone, asymmetric H-O-H combination band and Mg-OH combination bands, respectively. The band minimum ranges from 2.30 to 2.34 µm, suggesting a complex Fe/Mg composition of serpentine-group phyllosilicate. Generally, our results show that the absorption depth of the 0.7 µm feature is positively correlated with the absorption depth of the 2.3 µm feature. NWA 8534, NWA 3340 and QUE 93005 show strong absorption around 0.7 µm and 2.3 µm (Table 3), indicating they experienced a relatively high degree of aqueous alteration, which is consistent with ATR and XRD results. MCY 05231, NWA 8534, Murchison, QUE 93005, LAP 02333 and NWA 3340 show absorption features around 1.2 µm, suggesting the presence of FeII materials, such as olivine and phyllosilicate, while MIL 13005 shows an unusual 1.4 µm absorption, which might be caused by dehydration of the phyllosilicates. Y-86695 and B-7904 have no absorption features in the SWIR region and a reddened slope, which might represent a higher degree of thermal metamorphism to the point at which phyllosilicates recrystallise, although we didn't see any evidence for recrystallised phyllosilicates and secondary olivine in Y-86695. B-7904 did experience strong thermal metamorphism as suggested by SWIR, lacking an absorption feature at 2.3 µm, but with its 0.7 µm feature still preserved and shifted to shorter wavelengths, although the possibility of terrestrial weathering also could not be ruled out in this transition. According to Cloutis et al. (2012), B-7904 exhibits an absorption feature near 0.67 µm, suggesting the presence of saponite or olivine which may be caused by the presence of low Fe content in B-7904 (Cloutis et al. 2012). The high signal-to-noise ratio of the spectra in this study enables the detection of a weak 0.9 µm feature that confirms the occurrence of olivine.

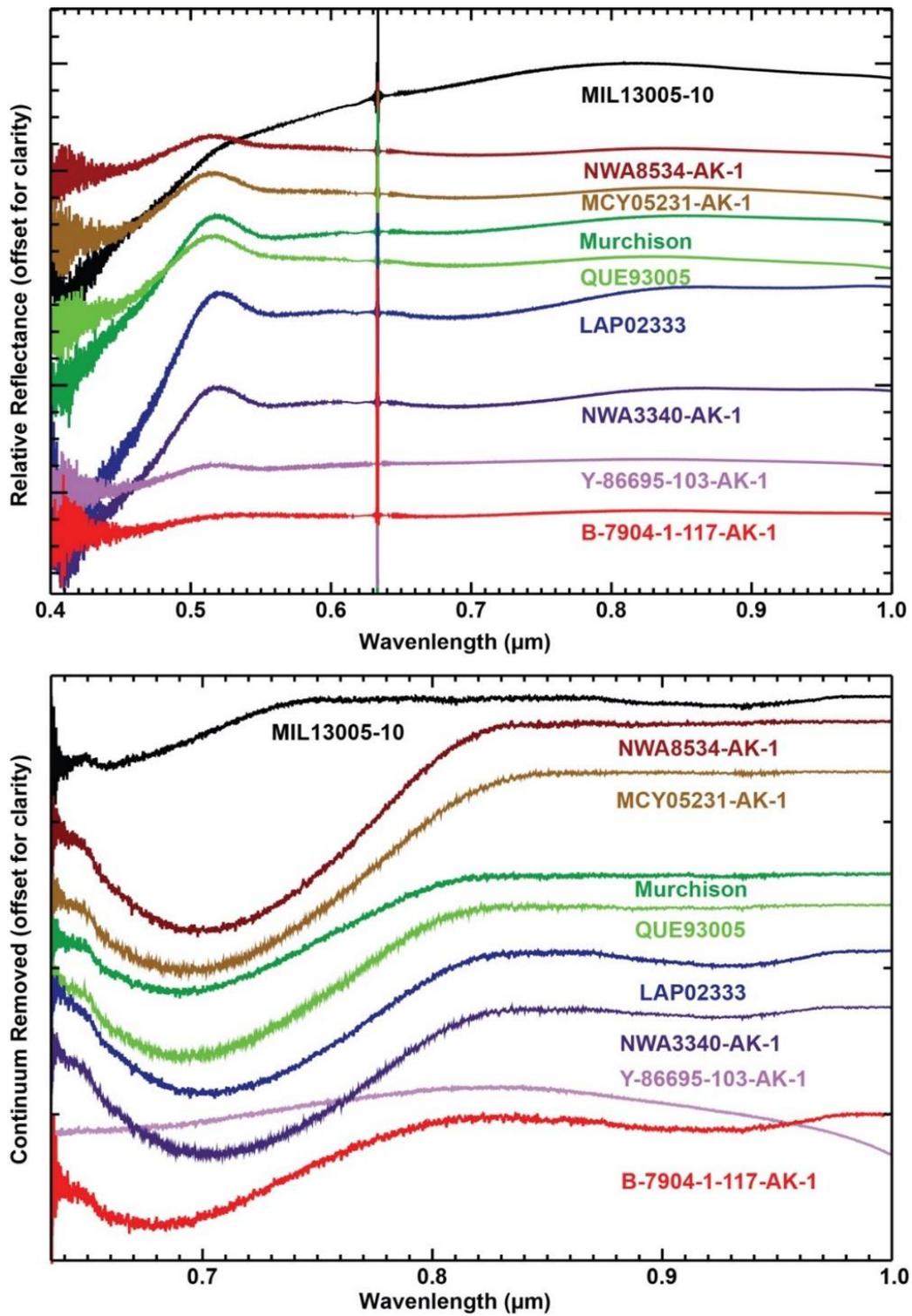

Figure 8. The upper panel shows visible/near infrared reflectance spectra between 0.4 and 1 μm for the aqueously altered CM samples and the lower panel shows enlarged view of continuum removed spectra between the range of 0.635 – 1 μm.

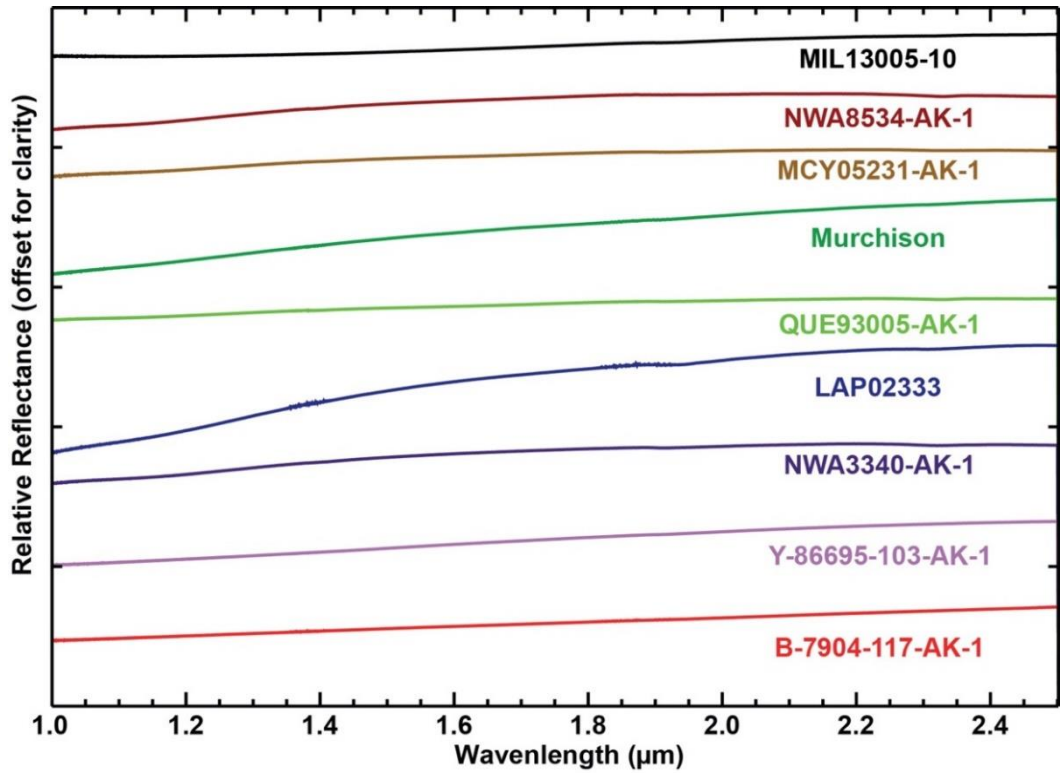

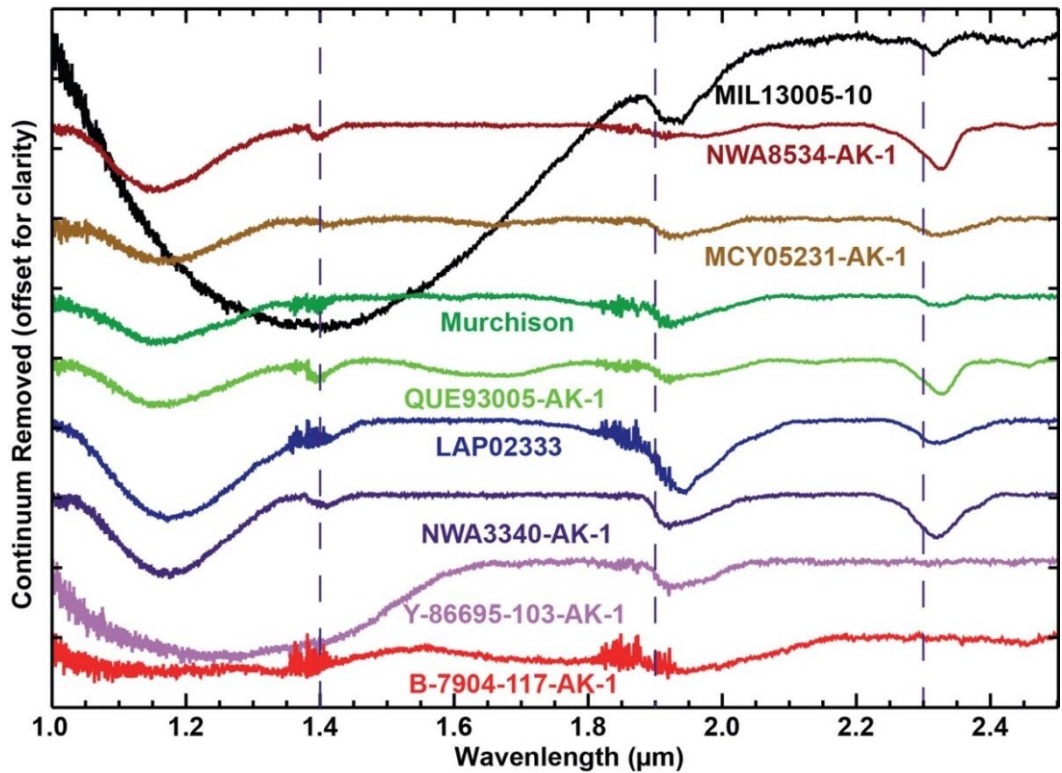

Figure 9. The upper panel presents shortwave infrared reflectance spectra in the range of 1 – 2.5 μm of CM samples while the lower panel shows continuum removed spectra.

Table 3. The Near Infrared (NIR) Spectral Parameters

| Meteorite | 0.700 μm feature | BD700 (%) | 2.3 μm feature | BD2300 (%) |
| --- | --- | --- | --- | --- |
| MIL13005-10 | - | - | 2.30 | 0.19 |
| Y-86695-103-AK-1 | - | - | - | - |
| NWA8534-AK-1 | 0.706 | 3.57 | 2.33 | 0.54 |
| NWA3340-AK-1 | 0.705 | 2.98 | 2.33 | 0.47 |
| MCY05231-AK-1 | 0.703 | 2.95 | 2.34 | 0.17 |
| LAP02333 | 0.704 | 2.91 | 2.34 | 0.24 |
| QUE93005-AK-1 | 0.700 | 2.63 | 2.33 | 0.41 |
| Murchison | 0.697 | 2.0 | 2.33 | 0.32 |
| B-7904-117-AK-1 | 0.687 | 1.80 | - | - |

## 3.4. Formation and evolution of CM parent body

The formation history of CM parent bodies in the early stages of the Solar System and the nature and extent of CM (and similar) chondrite alteration are not yet completely understood (Tomeoka et al. 1989; Bischoff 1998; Visser et al. 2018; Trigo-Rodríguez et al. 2019; Suttle et al. 2021). The mineralogy of a meteorite can provide important clues about the conditions under which it formed and the subsequent history of alteration it has undergone, including thermal and aqueous alteration events in the CM parent bodies.

Analysis of the CM chondrites in this study reveals the presence of phases such as phyllosilicates, magnetite, and carbonates that formed during water-rock reactions on the parent body's surface or within its interior (McSween Jr, H.Y. 1987; Brearley 2006). In most of the samples, olivine and pyroxene represent materials that were not consumed during the reactions. The degree and extent of aqueous alteration can vary depending on factors such as the duration, intensity of the alteration process, water/rock ratios, the primary mineralogy of the meteorite, and the chemistry of the fluids involved (Rubin 1997). Aqueous alteration significantly modified the original mineralogy of CM chondrites (McSween Jr, H.Y. 1999), which are typically brecciated and consist of clasts with various degrees of aqueous alteration (Metzler 1995; Bischoff et al. 2006; 2017) due to impacts that might have destroyed precursor planetesimals (Metzler et al. 1992; Bischoff 1998).

ATR and XRD results indicate that NWA 8534, MCY 05231, QUE 93005, NWA 3340 and Murchison experienced varying degrees of aqueous alteration, in good agreement with previous studies of these meteorites (e.g., King et al. 2017; 2021). XRD in particular indicates the presence of relatively crystalline phyllosilicates, which suggests that maximum temperature didn't reach > 300 °C after aqueous alteration had ended. This is also supported by the Raman characteristics, which are not consistent with significant heating (Fig. 6). NIR reflectance spectra of Murchison and MCY 05123 detect the presence of phyllosilicate and is corresponded to the published literature (Bates et al. 2020; Kiryu et al. 2020).

D and G bands in the Raman spectra of NWA 3340 and LAP 02333 reveal that they were subjected to lower temperatures (150-200 °C), which are the estimated alteration temperatures for CM chondrites. It is likely that long-term heating occurred in their parent bodies (Busemann et al. 2007). Presence of goethite in LAP 02333 reveals that the meteorite had experienced terrestrial weathering. Identification of diopside by Raman spectroscopy reveals that the LAP 02333 contains unaltered chondrules and CAIs. XRD data confirm the presence of serpentine in LAP 02333, as it wasn't heated above 200 – 300 °C. NIR-VIS spectroscopy results detect phyllosilicates in the sample. NWA 3340 is dominated by serpentine and very low percentage of olivine based on our XRD data.

The XRD pattern for MIL 13005 indicates that it is dominated by magnetite and iron sulphide and lacks crystalline serpentine. The XRD pattern is typical of CM chondrites that experienced short-lived thermal metamorphism at ~300 – 500 °C, temperatures high enough to partially dehydrate phyllosilicates and cause them to lose their crystalline structure (e.g., Nakamura 2005; King et al. 2021). We note that this first report of thermal metamorphism for MIL 13005 is supported by its distinct IR spectra. However, the Raman spectral features of MIL 13005 show no clear evidence for heating, although this is perhaps not surprising given the study of Kiryu et al. (2020) showing that temperatures of >600°C are required to significantly modify the structure of organic materials on short timescales.

XRD results show that Y-86695 also lacks crystalline phyllosilicates due to thermal metamorphism. It contains a relatively high abundance of olivine, which based on the sharp peaks in the XRD pattern is likely primary material that survived aqueous alteration, rather than secondary olivine recrystallized during metamorphism. This constrains the peak metamorphic temperature for Y-86695 to <500°C (e.g., Nakamura 2005), in agreement with the study of King et al. (2021) and consistent with Raman spectral properties that cannot be distinguished from strongly heated CM chondrites (e.g., Quirico et al. 2018, Kiryu et al. 2020). In general, the spectra for Y-86695 and MIL 13005 differ from most samples, which is consistent with it being heated. Identification of iron oxide (goethite) by Raman spectroscopy in Y-86695 as secondary mineral is probably caused by terrestrial weathering. Raman spectra of feldspar in Y-86695 could be associated with chondrules and CAIs, as this mineral is rare in CM chondrites. These findings are consistent with the results in the literature (Nakamura 2005; King et al. 2021).

In comparison, most of the olivine in B-7904 is Fe-rich and poorly crystalline (the broad peaks on Fig. 1) because it formed during metamorphism at temperatures > 500 °C. Spectroscopy and ATR results also detect recrystallized phyllosilicate in B-7904, supporting the higher peak metamorphic temperature. The Raman spectral data (Fig. 6) and higher intensity of D and G band compared to other samples (Fig. 7) further support B-7904 being strongly heated, which is consistent with previous studies and current XRD/IR results .According to the previous studies, B-7904 experienced heating at a temperature higher than 700 °C and no hydrous minerals have been detected (Nakamura 2005; Nakato et al. 2008; King et al. 2021).

MCY 05231 and Murchison show no evidence of heating and thermal metamorphic history based on our results from Raman spectroscopy, ATR and XRD, which is also consistent with the result from previous studies (King et al. 2017; Bates et al. 2020; King et al. 2021). NIR reflectance spectra of Murchison and MCY 05123 detect the presence of phyllosilicate and is corresponded to the published literature (Bates et al. 2020; Kiryu et al. 2020).

Phosphates are another clue to lead us to discover more about the evolution of asteroid parent bodies. Phosphate is a secondary mineral in carbonaceous chondrites which might have formed during parent body hydrothermal processing (Flynn et al. 2022). In a broad sense, phosphate minerals are generally more common in CI chondrites, with phosphates being rare in CM chondrites (Brearley et al. 2005). However, there are some reports of phosphate occurrences in CM chondrites in the literature (Brearley et al. 2005; Nazarov et al. 2009; Singerling et al. 2018). The presence of phosphate minerals in CM chondrites may suggest that these meteorites experienced similar aqueous alteration processes as the CI chondrites. Aqueous alteration likely happened in the presence of water-rich fluids, containing cations necessary for phosphate mineral formation such as such as calcium, magnesium, or iron. Furthermore, phosphorus-bearing minerals, such as apatite, whould have reacted with the fluid during the alteration process. Singerling et al. (2018) suggested that the formation of phosphates during aqueous alteration in CMs also required suitable temperature and pressure conditions. Generally, moderate to low temperatures and pressures are conducive to phosphate formation (Singerling et al. 2018). The identification of phosphate in MIL 13005 expands the known mineral diversity of CM chondrites, broadening our understanding of their mineralogy and providing new insights into their formation processes.

In summary, phosphates are important for understanding conditions on the CM parent body because they provide insights into aqueous alteration processes, temperature and pressure conditions, redox conditions, chronology, and the origin of water and volatile elements. Studying phosphates in CM chondrites can contribute significantly to our understanding of the parent body's formation and evolution, but also offers insights into broader questions regarding the early solar system, the habitability of celestial bodies, and the potential for life beyond Earth.

## 4. Conclusions

(I) CM carbonaceous chondrites originate from primitive water-rich asteroids formed during the early solar system and they likely delivered abundant organic building blocks of life required for biogenesis on early Earth and Mars. We have used Raman spectroscopy to determine the composition and mineralogy of eight CM and one ungrouped carbonaceous chondrite. Visible to Near Infrared and ATR spectroscopy and XRD were also used to characterize the mineralogy and aqueous and thermal alteration history of the samples. In summary, we find that: The presence of the D-band in Raman spectra of all samples shows that carbon content in these samples is highly disordered in nature. The FWHM of D-band is the characteristic of D-band and is used to estimate the highly metamorphosed CM chondrites in this study.

(II) The extent of thermal metamorphism in the chondrites investigated in this study is as follows; B-7904 is a stage IV that was heated to high enough temperatures to recrystallize the phyllosilicates back into olivine and pyroxene. MIL 13005 and Y-86695 are stage II heated CMs, where thermal metamorphism led to the dehydration of phyllosilicates. LAP 02333 looks like an unheated CM but has some unusual Raman and IR properties. NWA 8534, MCY 05231, QUE 93005, Murchison, and NWA 3340 are unheated CM chondrites that experienced varying degrees of aqueous alteration.

(III) The VNIR and ATR spectra of B-7904 (C2-ung) exhibits a weak 0.7 µm absorption that can be attributed to the mixed valence $Fe^{2+}$-$Fe^{3+}$ phyllosilicate present in the NIR wavelength region. Although the sample contains secondary olivine, the lack of distinct olivine absorption bands implies that thermal metamorphism did not lead to the formation of widespread crystalline Fe2+-bearing olivine, which would have otherwise aided in differentiating between primary and recrystallized olivine. The SWIR region of the spectrum appears flat and featureless, while the MIR region displays characteristic olivine Si-O stretching and bending. These observations suggest that the sample has undergone a higher degree of thermal metamorphism and its phyllosilicate has been recrystallized, which is consistent with the XRD data and previous studies.

(VI) Raman spectroscopy data, in addition to VIS-NIR spectra, could be used to infer geological processes on asteroid surfaces. By combining Raman spectroscopy and VIS-NIR data of chondrites, we can obtain a more comprehensive understanding of the geological processes that occurred on their parent asteroid surfaces such as thermal history, degree of shock metamorphism, and possible impact events. These data sets provide complementary information on the mineralogy, thermal history, shock metamorphism, aqueous alteration, and regolith evolution of the asteroid, helping us better understand the early solar system's history and conditions.

(VII) The absorption band found on Bennu's surface and in CM chondrites, indicating a similar aqueous history and suggesting a common formation history and aqueous alteration processes in the early solar system. The reflectance spectra of LAP 02333 and Bennu has similar infrared spectral features in the range of MIR region (Hamilton et al. 2019).

(VIII) The detection of phosphate in the CM chondrite MIL 13005 demonstrates that phosphate minerals are not exclusive to CI chondrites. This finding broadens our understanding of the mineralogy of CM chondrites and provides new insights into their formation processes. Detailed mineralogical and petrographic analysis could be done by conducting further in-depth analyses of phosphate minerals in CM chondrites using advanced techniques such as scanning electron microscopy (SEM) or transmission electron microscopy (TEM) to better understand their morphology, composition, and distribution within the meteorite. For a broader understanding of CM chondrites and their formation processes, one potential future study that could build upon the findings of phosphate minerals in CM chondrites could be investigating the relationship between phosphate minerals and organic matter. Since CM chondrites are known to contain organic matter, it would be interesting to explore the potential interactions between phosphate minerals and organic matter within the meteorites, and their implications for understanding the early solar system.

**Acknowledgements:** AK acknowledges funding support from UK Research and Innovation (UKRI) grant MR/T020261/1. The National Institute of Polar Research (NIPR), Japan, is thanked for providing the samples of Y-86695 and B-7904. US Antarctic meteorite samples are recovered by the Antarctic Search for Meteorites (ANSMET) program which has been funded by NSF and NASA, and characterised and curated by the Department of Mineral Sciences of the Smithsonian Institution and Astromaterials Curation Office at NASA Johnson Space Center. Hot desert meteorite samples were provided by Laurence Garvie at the Center for Meteorite Studies, Arizona State University, USA.